\documentclass[10pt,conference]{IEEEtran}
\usepackage{array}
\usepackage{cite}
\usepackage{calc}
\usepackage{caption}
\usepackage{graphicx}
\usepackage{psfrag}
\usepackage[tight,footnotesize]{subfigure}
\usepackage{stfloats}
\usepackage{url}
\usepackage{subfig}
\usepackage{longtable}
\usepackage{stfloats}
\usepackage{amsfonts,amssymb,amsbsy,bm,paralist,theorem,cite,ifthen,color}
\usepackage[cmex10]{amsmath}
\usepackage{algorithmic,algorithm}

\newcommand\eb{\ensuremath{{\bm e}}}
\newcommand\wb{\ensuremath{{\bm w}}}

\newcommand\hb{\ensuremath{{\bm h}}}
\newcommand\Ab{\ensuremath{{\bm A}}}

\newcommand\Ib{\ensuremath{{\bm I}}}

\newcommand\Yb{\ensuremath{{\bm Y}}}

\newcommand\Rb{\ensuremath{{\bm R}}}

\newcommand\Vb{\ensuremath{{\bm V}}}
\newcommand\Vc{\ensuremath{{\mathcal{V}}}}

\newcommand\Wb{\ensuremath{{\bm W}}}

\newcommand\zerob{\ensuremath{{\bm 0}}}

\newcommand\Psib{\ensuremath{{\bf \Psi}}}

\newcommand\tr{\ensuremath{{\rm Tr}}}

\newcommand\rank{\ensuremath{{\rm rank}}}

\newcommand\SINR{\ensuremath{{\sf SINR}}}

\newcommand\Cplx{\ensuremath{{\mathbb{C}}}}
\newcommand\Ral{\ensuremath{{\mathbb{R}}}}

\newtheorem{lemma}{Lemma}
\newtheorem{proposition}{Proposition}

\newtheorem{Fact}{Fact}

\newtheorem{Condition}{Condition}
\newtheorem{Claim}{Claim}

\begin{document}
\bibliographystyle{IEEEtran}
\title{Worst-Case Robust Multiuser Transmit Beamforming Using Semidefinite Relaxation: Duality and Implications}

\author{Tsung-Hui Chang$^\star$, Wing-Kin
Ma$^\dag$, and Chong-Yung Chi$^\star$ \\ ~ \\
\begin{tabular}{cc}
$^\star$Institute of Commun. Eng. \& Dept. of Elect. Eng.   &
$^\dag$Department of Electronic Engineering   \\
National Tsing Hua University,   &    The Chinese University of Hong Kong, \\
Hsinchu, Taiwan 30013     &    Shatin, N.T.,  Hong Kong \\
\small E-mail: tsunghui.chang@ieee.org, cychi@ee.nthu.edu.tw & \small E-mail:
wkma@ieee.org
\end{tabular}
\vspace{-\baselineskip}}

\maketitle
\begin{abstract}
This paper studies a downlink multiuser transmit beamforming design under spherical channel uncertainties, using a worst-case robust formulation.
This robust design problem is nonconvex.
Recently, a convex approximation formulation based on semidefinite relaxation (SDR) has been proposed to handle the problem.
Curiously, simulation results have consistently indicated
that SDR can attain the global optimum of the robust design problem.
This paper intends to provide some theoretical insights into this important empirical finding.
Our main result is a dual representation of the SDR formulation,
which reveals an interesting linkage to a different robust design problem, and the possibility of SDR optimality.
\end{abstract}

\IEEEpeerreviewmaketitle
\vspace{-0.2cm}
\section{Introduction}
%Multiple-antenna transmission has been a promising approach to
%obtaining significant throughput gains in wireless broadcast
%channels \cite{Goldsmith2003}. In view of the complexity of dirty
%paper coding, linear transmit beamforming has been recognized as a
%powerful transmission technique since it can achieve a large
%fraction of capacity with a much lower implementation complexity
%\cite{SharifHassibi2007}.
This paper focuses on a standard wireless multiuser unicast
system where a multiple-antenna transmitter broadcasts independent
data streams to multiple single-antenna receivers using transmit
beamforming \cite{BK:Bengtsson01}. %The beamforming vectors are designed to fulfill the quality-of-service
%(QoS) requirements of receivers. %, the transmitter aims to form
%appropriate beam patterns to enhance the strength of
%signal-of-interest for each of the receivers while mitigating the
%co-channel interference among the receivers at the same time.
%Transmit beamforming designs, especially those that use the
%signal-to-interference-plus-noise ratio (SINR) as the receivers' QoS
%measure, are challenging because they often involve nonconvex
%optimization problems, which are known difficult in general. Recent
%developments have shown that the unicast transmit beamforming
%problem can actually be solved efficiently through proper problem
%reformulation and convex optimization theory; see
%\cite{BK:Bengtsson01,WieselTSP06} and also
%\cite{Gershman2010_SPM} for a comprehensive review.
In this context, the efficacy of beamforming designs
relies on knowledge of the channel state information (CSI) of all
the receivers.
However, the transmitter often has some uncertainties on the CSI, due to
issues such as finite-length training and finite-rate feedback \cite{Love2008}.
CSI uncertainties at the transmitter can result in significant performance outage,
if not taken into consideration in the beamforming designs.
%Under such circumstances,
%receivers may suffer from performance outage.
%Such
%CSI-uncertainty-induced performance loss
The CSI uncertainty problem
has motivated considerable research endeavors in
robust transmit beamforming design techniques.
%that take into consideration the CSI uncertainties.
This includes
the
chance constrained robust designs
%\cite{Shenouda2008,Vucic2009_chanceMSE},
\cite{Shenouda2008,our_outage_SDR},
where the CSI uncertainties are modeled as
random variables, and the worst-case robust designs
\cite{Shenouda2009_WorstcaseMSE,Shenouda2007,Boche2009,ZhengWongNg_2008}, where the CSI uncertainties are modeled as bounded unknowns
within a predetermined, small error set. %In the former design, the beamforming vectors are designed such that
%all the receivers can achieve the QoS requirements with a high
%probability (i.e., a low performance outage probability); while, in the latter design, the beamforming vectors
%are designed such that QoS requirements are achieved for all
%channels satisfying this bounded error model.

%Our interest in this paper lies in a
Our problem of interest is the
worst-case signal-to-interference-plus-noise ratio (SINR) constrained robust
transmit beamforming design problem under spherically bounded CSI uncertainties,
which has drawn much interest recently
\cite{Shenouda2007,Boche2009,ZhengWongNg_2008}.
%Here SINR is used as a measure of the receiver performance.
Presently available beamforming solutions for this worst-case robust problem are based on approximation methods,
either restriction~\cite{Shenouda2007,Boche2009} or relaxation~\cite{ZhengWongNg_2008},
and it is now not clear whether the worst-case robust problem can be optimally (and efficiently) solved.
However, simulations seem to have provided the answer to the latter--- the semidefinite relaxation (SDR) method~\cite{ZhengWongNg_2008}.
SDR is a convex relaxation technique for a certain class of hard (nonconvex) optimization problems,
and has recently gained popularity owing to its wide scope of applicability~\cite{BK:LuoChang,Luo2010_SPM}.
For a general application, SDR is considered a suboptimal solver;
however, for the worst-case robust beamforming problem,
simulation results have indicated that SDR {\it should} to be a globally optimal solver, which is a rather surprising empirical finding.
As such, being able to provide
a theoretical analysis proving whether SDR is optimal % (or counter-proving it)
would be of much significance.
A recent result~\cite{Song2011} has partially addressed this open question,
where the SDR optimality under sufficiently small error radii is analyzed.

This paper intends to address the mystery of SDR optimality in worst-case robust transmit beamforming optimization using a different analysis approach.
We show that the worst-case robust problem has a close relationship to a different robust beamforming problem, in form of max-min optimization.
In particular, we prove that their SDR problems are dual, or equivalent, to each other.
This new, intriguing, duality relationship provides a new perspective and useful insights explaining the optimality of SDR.
In particular,
we will give a condition under which SDR provides globally optimal solutions to the worst-case robust problem.

%%%%%%%%%%%%%%%%%%%%%%%%%%%%%%%%%%%%%%%%%%%%%%%%%%%%%%%%%%%%%%%%%%%%%%%%
\section{Signal Model and Background}
%%%%%%%%%%%%%%%%%%%%%%%%%%%%%%%%%%%%%%%%%%%%%%%%%%%%%%%%%%%%%%%%%%%%%%%%
%\vspace{-0.1cm}
%%%%%%%%%%%%%%%%%%%%%%%%%%%%%%%%%%%%%%%%%%%%%%%%%%%%%%%%%%%%%%%%%%%%%%%%%
%\subsection{Signal Model and Problem Statement}
%%%%%%%%%%%%%%%%%%%%%%%%%%%%%%%%%%%%%%%%%%%%%%%%%%%%%%%%%%%%%%%%%%%%%%%%%
\vspace{-0.1cm}
Consider a wireless downlink system where a transmitter,
equipped with $N_t$ antennas, wants to communicate with
$K$ single-antenna receivers using transmit beamforming.
The problem formulation follows a standard unicast setting~\cite{BK:Bengtsson01}:
Let $\hb_i\in\mathbb{C}^{N_t}$ denote the channel vector of
receiver $i$, and let $\wb_i\in\mathbb{C}^{N_t}$ be the
associated beamforming vector for receiver $i$.
The SINR of receiver $i$
is given by
\begin{align}
{\SINR}_i(\wb_1,\ldots,\wb_K,\hb_i)  =\frac{|\hb^H_i \wb_i|^2}{\sum_{k\neq i}^{K}
|\hb_i^H\wb_k|^2+\sigma_i^2},
\end{align}where $\sigma_i^2>0$ is the noise power at receiver $i$, for all $i=1,\ldots,K$.
Our goal is to design the beamforming vectors $\{\wb_i\}_{i=1}^K$
such that each receiver achieves a desired SINR level.

Conventionally, transmit beamforming designs
require full channel state information (CSI) at the transmitter; i.e.,
knowledge of $\{\hb_i\}_{i=1}^K$.
In wireless communications, however, it is difficult for the transmitter to acquire accurate CSI, due to imperfect channel estimation and finite rate feedback \cite{Love2008}. Hence there are channel uncertainties at the transmitter; i.e.,
\begin{align}
 \hb_i= \bar\hb_i+\eb_i,~i=1,\ldots,K, \label{channel model}
\end{align}where $\bar\hb_i$ denotes the channel estimate available
at the transmitter, and
$\eb_i  \in \Cplx^{N_t}$ represents the channel uncertainty. In this work, we focus on spherically bounded channel uncertainties:
% Specifically, we assume
%that each $\eb_i$ lies in a Euclidean norm ball:
\begin{align}\label{EC}
  \|\eb_i\|^2 \leq r_i^2,~i=1,\ldots,K,
\end{align}where $\| \cdot \|$ denotes the Euclidean norm, and
$r_i>0$ is the radius of the uncertainty ball. %Conventional beamforming designs that do not take these channel uncertainties into consideration cannot provide guaranteed performance for the receivers.
We study the following worst-case robust beamforming design \cite{ZhengWongNg_2008,Shenouda2007}:
\begin{subequations} \label{eq. robust beamforming}
\begin{align}
\!\!\!\!\min_{\substack{\wb_i \in \mathbb{C}^{N_t}, \\ i=1,\ldots,K}}~ &\sum_{i=1}^{K} \|\wb_i\|^2\\
{\rm s.t.} &~ {\SINR}_i(\wb_1,\ldots,\wb_K,\bar\hb_i+\eb_i) \geq \gamma_i~\forall~\|\eb_i\|^2 \leq r_i^2,\notag \\
&~i=1,\ldots,K, \label{eq. robust beamforming C1}
\end{align}
\end{subequations} where $\gamma_i>0$ is the SINR requirement of receiver $i$,
which must be fulfilled even under worst possible CSI uncertainties.
%Formulation \eqref{eq. robust beamforming} aims to provide guaranteed SINR performance for the receivers.

The challenge of solving the worst-case robust problem \eqref{eq. robust beamforming} lies in the worst-case SINR constraints in \eqref{eq. robust beamforming C1},
each of which corresponds to an infinite number of nonconvex quadratic constraints.
As mentioned, there are several approximation methods for managing problem \eqref{eq. robust beamforming} \cite{ZhengWongNg_2008,Shenouda2007,Boche2009},
and here we focus on the SDR method~\cite{ZhengWongNg_2008}.
The development of SDR consists of two steps.
The first step, which is standard (see, e.g. \cite{Luo2010_SPM}),
is to substitute $\Wb_i = \wb_i \wb_i^H$, $k=1,\ldots,K$, into \eqref{eq. robust beamforming C1},
and then replace $\Wb_i = \wb_i \wb_i^H$ by $\Wb_i \succeq {\bf 0}$ (i.e., $\Wb_i$ being positive semidefinite (PSD)) to obtain a relaxed problem
\begin{subequations} \label{eq. robust beamforming2}
\begin{align}
\!\!\!\!\min_{\substack{\Wb_i \in \mathbb{H}^{N_t}, \\ i=1,\ldots,K}}~ &\sum_{i=1}^{K} \tr(\Wb_i) \\
{\rm s.t.} &~
( \bar\hb_i+ \eb_i )^H \left(\frac{1}{\gamma_i}\Wb_i-\sum_{k\neq
i}^{K}\Wb_k\right) ( \bar\hb_i+ \eb_i ) \geq \sigma_i^2 \notag \\
&~ \forall ~\|\eb_i\|^2 \leq r_i^2, \quad k=1,\ldots,K, \label{eq. robust beamforming2 C1} \\
&~\Wb_1,\ldots,\Wb_K \succeq {\bf 0},
\end{align}
\end{subequations}
where $\mathbb{H}^{N_t}$ is the set of all $N_t$ by $N_t$ Hermitian matrices, and
${\tr}(\Wb_i)$ denotes the trace of $\Wb_i$.
The motivation of this step is to linearize the nonconvex constraints.
The second step is to turn \eqref{eq. robust beamforming2 C1} to finite numbers of constraints,
thereby enabling efficient implementations.
By applying $\mathcal{S}$-lemma (see \cite{BK:BoydV04}) to \eqref{eq. robust beamforming2 C1},
we obtain the following SDR formulation of \eqref{eq. robust beamforming}:
\begin{subequations}\label{WSP-SDR}
\begin{align}
\!\!\!\!\!\!\min_{\substack{\Wb_i
\in \mathbb{H}^{N_t}, \lambda_i\in \Ral, \\ i=1\ldots,K}}~
&\sum_{i=1}^{K} {\tr}(\Wb_i) \\
{\rm s.t.}~ &
\Psib_i\left(\Wb_1,\ldots,\Wb_K,\lambda_i\right)\succeq
{\bf
0},~i=1,\ldots,K, \label{WSP-SDR C1}\\
&\Wb_1,\ldots,\Wb_K \succeq {\bf 0},~\lambda_1,\ldots,\lambda_K\geq
0,\notag
\end{align}
\end{subequations}
where the matrix functions $\Psib_i\left(\Wb_1,\ldots,\Wb_K,\lambda_i\right)$ are defined as
\begin{align} \label{definition of Psib_i}
&\Psib_i\left(\Wb_1,\ldots,\Wb_K,\lambda_i\right)\triangleq
\begin{bmatrix}
\Ib \\
\bar \hb_i^H
\end{bmatrix}\left(\frac{1}{\gamma_i}\Wb_i-\sum_{k\neq
i}^{K}\Wb_k\right)
\begin{bmatrix}
\Ib ~\bar \hb_i
\end{bmatrix}\notag \\
&~~~~~~~~~~~~~~~~~~~+
\begin{bmatrix} \lambda_i \Ib & {\bf 0}\\
{\bf 0} & -\sigma_i^2-\lambda_ir_i^2
\end{bmatrix},~i=1,\ldots,K,
\end{align} where $\Ib$ is the $N_t$ by $N_t$ identity matrix.
Note that the SDR problem \eqref{WSP-SDR} is a semidefinite program (SDP), which is convex and tractable.
%$\Ib$ is the $N_t$ by $N_t$ identity matrix, ${\tr}(\Wb_i)$ denotes the trace of $\Wb_i$, and
%$\Wb_i\succeq \zerob$ means that $\Wb_i$ is positive semidefinite (PSD).

The SDR problem \eqref{WSP-SDR} is methodologically an approximation to the worst-case robust problem \eqref{eq. robust beamforming}
because the ranks of $\Wb_i$ are not constrained.
However, if the optimal solution of the SDR problem \eqref{WSP-SDR},
denoted by $( \Wb_1^\star, \ldots, \Wb_K^\star )$,
is of rank one; i.e., $\Wb_i^\star=\wb_i^\star(\wb_i^\star)^H$ for all $i=1,\ldots,K$,
then it can be verified that $(\wb_1^\star,\ldots,\wb_K^\star)$ is a globally optimal solution to the worst-case robust formulation \eqref{eq. robust beamforming}.
Rather surprisingly, it is found through simulations that
SDR yields rank-one solution automatically, and it happens seemingly all the time  \cite{ZhengWongNg_2008,Song2011} (see also \cite{our_outage_SDR}).
Our endeavor in the subsequent section is to provide a dual formulation of the SDR problem \eqref{WSP-SDR} that may shed light into this empirical finding.

%which links SDR to a different robust design problem. The linked robust design problem
%enriches our fundamental understanding of the original problem \eqref{eq. robust beamforming},
%and may shed light into the optimality of its SDR problem \eqref{WSP-SDR}.

%
%some clues on the rank-one optimality of SDR, which may help enrich our fundamental understanding of the worst-case robust beamforming design problem.
%We will present a dual formulation of the SDR problem \eqref{WSP-SDR}
%which links SDR to a different robust design problem
%and may shed light into the optimality of SDR to the original problem \eqref{eq. robust beamforming}.

Before we proceed to the main result,
%we
let us present some simulation results to further strengthen the motivation of the raised analysis problem.
Specifically, we benchmark the SDR method against other concurrent approximation methods, namely, the robust SOCP-based method in
\cite{Shenouda2007}, and the MMSE-based SDP method in \cite{Boche2009}.
The simulation settings are: $N_t= 4$, $K= 4$,
$\gamma \triangleq \gamma_1 = \cdots=
\gamma_K,$
$\sigma_1^2 =\cdots=\sigma_K^2=0.1,$
$r\triangleq r_1=\cdots=r_K=0.1$,
and $(\bar\hb_1,\ldots,\bar\hb_K)$
being independent and identically distributed complex Gaussian random variables with zero mean and unit variance.
The result is shown in Fig.~1, where
we see that the SDR method outperforms the other two methods.
Moreover, we should emphasize that the SDR method yielded rank-one solution in all the trials ran.

\vspace{-0.2cm}
%%%%%%%%%%%%%%%%%%%%%%%%%%%%%%%%%%%%%%%%%%%%%%%%%%%%%%%%%%%%%%%%%%%%%%%%
%\section{A Dual Formulation of \eqref{WSP-SDR}}
\section{Duality of Worst-Case Robust SDR}
%%%%%%%%%%%%%%%%%%%%%%%%%%%%%%%%%%%%%%%%%%%%%%%%%%%%%%%%%%%%%%%%%%%%%%%%
\vspace{-0.1cm}

Consider the following max-min optimization problem
\begin{align}\label{WPP}
\max_{\substack{\eb_i\in \Cplx^{N_t}, \\
i=1,\ldots,K}}~ &\left\{
 \begin{array}{ll}
  {\displaystyle \min_{\substack{\wb_i \in \mathbb{C}^{N_t}, \\ i=1,\ldots,K}}}~ &{\displaystyle \sum_{i=1}^{K} ||\wb_i||^2}\\
  ~~~~~{\rm s.t.}&~ \SINR_i(\wb_1,\ldots,\wb_K,\bar\hb_i+\eb_i) \geq
  \gamma_i,\\
  &~i=1,\ldots,K,
 \end{array}\!\!\!\!\right\} \notag \\
{\rm s.t.}~&~ \|\eb_i\|^2\leq r_i^2,~  i=1,\ldots,K.
\end{align}
At first look, problem~\eqref{WPP} is different from the worst-case robust problem in \eqref{eq. robust beamforming}.
In \eqref{WPP},
the inner minimization %part
is a standard non-robust beamforming design problem \cite{BK:Bengtsson01} which finds the most power efficient design given a presumed CSI $\{\bar\hb_i+\eb_i\}_{i=1}^K$.
%, for a particular CSI uncertainty $\{\eb_i\}_{i=1}^K$.
The outer maximization, %part,
however, targets to find a ``worst'' set of CSI uncertainties $\{\eb_i\}_{i=1}^K$ that maximizes
the inner-minimum transmit power.
%the required transmission power for the associated inner beamforming problem.
%Problem \eqref{WPP} can also be interpreted as a zero-sum game problem between the two players of the CSI error vectors $\{\eb_i\}$ and the beamforming vectors $\{\wb_i\}$.
We should also note that problem \eqref{WPP} has a flavor of two-player zero-sum game.

\begin{figure}[!t] \centering
   \resizebox{0.40\textwidth}{!}{
     \psfrag{A}[Bl][Bl]{\huge Average transmission power (dB)}
     \psfrag{SDP method}[Bl][Bl]{\LARGE MMSE-based SDP~\cite{Boche2009}}
     \psfrag{Structured SDP method}[Bl][Bl]{\LARGE Robust SOCP via structured SDP \cite{Shenouda2007}}
     \psfrag{SDR method in (5)}[Bl][Bl]{\LARGE SDR in (5)}
     %\psfrag{R1}[Bl][Bl]{\LARGE \cite{Shenouda2007}}
%     \psfrag{R2}[Bl][Bl]{\LARGE \cite{Boche2009}}
     \psfrag{R1}[Bl][Bl]{\LARGE }
     \psfrag{R2}[Bl][Bl]{\LARGE }
     \psfrag{gamma}[Bl][Bl]{\huge $\gamma$ (dB)}
      \includegraphics{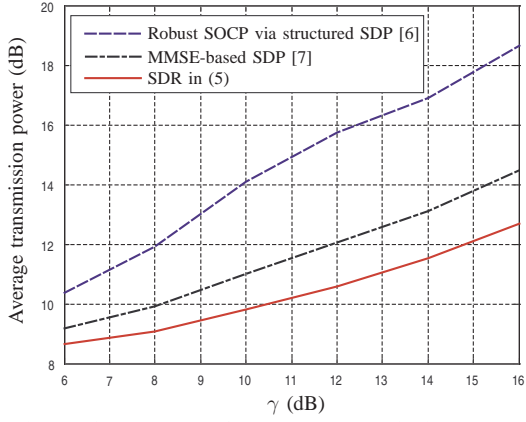}}
      \vspace{-0.3cm}
      \caption{Simulation results of average transmission power versus target SINR $\gamma$, for uncertainty radius $r=0.1$.}
     \label{fig:power_vs_gamma}\vspace{-0.5cm}
\end{figure}

We are particularly interested in applying SDR to \eqref{WPP}.
Like SDR for the worst-case robust problem,
we replace each $\wb_i\wb_i^H$ with a PSD matrix $\Wb_i$,
and each $[\eb_i^H~1][\eb_i^H~1]^H$ with a PSD matrix $\Vb_i$,
to obtain the following problem
 \begin{align}\label{WPP-SDR}
\max_{\substack{\Vb_i \in \mathbb{H}^{N_t+1},\\i=1,\ldots,K}}
&\left\{\!\!\!\!\!\!
 \begin{array}{ll}
    &\!\!\!\!{\displaystyle\min_{\Wb_i\in  \mathbb{H}^{N_t}}~ \sum_{i=1}^{K}
    {\tr}(\Wb_i)}\\
    &~~~~~~~{\rm s.t.}~
    {\displaystyle \tr\left(\left(\frac{1}{\gamma_i}\Wb_i - \sum_{k\neq
      i}^{K}\Wb_k\right)\Rb_i\right) \geq \sigma_i^2,} \\
      &~~~~~~~~~~~~~~i=1,\ldots,K, \\
      &~~~~~~~~~~~~~~\Wb_1,\ldots,\Wb_K \succeq \zerob.
  \end{array}\!\!\!\!\right\}  \notag \\
{\rm s.t.}&~\tr(\Vb_i) \leq (1+r_i^2),~i=1,\ldots,K, \notag \\
         &~[\Vb_i]_{N_t+1}=1,~i=1,\ldots,K, \notag \\
         &~\Vb_1,\ldots,\Vb_K \succeq \zerob,
\end{align}
where $[\Vb_i]_{N_t+1}$ is the $(N_t+1,N_t+1)$th entry of $\Vb_i$ and
$
    \Rb_i=\begin{bmatrix}
      \Ib~  \bar\hb_i
    \end{bmatrix}
    \Vb_i
    \begin{bmatrix}
      \Ib~  \bar\hb_i
    \end{bmatrix}^H,~i=1,\ldots,K.
$

An important observation of problem \eqref{WPP-SDR} is that there always exists a rank-one solution for the inner
minimization of problem \eqref{WPP-SDR}:\vspace{-0.3cm}
\begin{Fact}{\rm \cite{BK:Bengtsson01}}
 Consider the following SDP:
 \begin{align}\label{PSDR}
\min_{\substack{\Wb_i \in \mathbb{H}^{N_t},\\i=1,\ldots,K}}~
&\sum_{i=1}^{K} {\tr}(\Wb_i) \\
{\rm s.t.}~&
 \tr\left(\left(\frac{1}{\gamma_i}\Wb_i - \sum_{k\neq
  i}^{K}\Wb_k\right)\Rb_i\right) \geq \sigma_i^2, ~i=1,\ldots,K,\notag\\
&\Wb_i \succeq {\bf 0}, ~i=1,\ldots,K,\notag
\end{align} where $\Rb_1,\ldots,\Rb_K\succeq \zerob$. Suppose that \eqref{PSDR} is feasible. Then there exists an optimal solution $(\Wb_1^\star,\ldots,\Wb_K^\star)$ for
which $\rank(\Wb_i^\star)= 1$ for all $i$.
\end{Fact}
Fact 1 implies that the SDR of $(\Wb_1,\ldots,\Wb_K)$ is always tight for the max-min SDR problem \eqref{WPP-SDR}. Fact 1
raises an intriguing question---What is the relationship between the max-min SDR problem \eqref{WPP-SDR} and the robust SDR problem \eqref{WSP-SDR}? If the optimal solutions of $(\Wb_1,\ldots,\Wb_K)$ of the two problems are identical, then Fact 1 immediately implies that \eqref{WSP-SDR} has a rank-one optimal solution and hence SDR is tight to \eqref{WSP-SDR} as well.

\subsection{Main Result}

It turns out that problems \eqref{WPP-SDR} and \eqref{WSP-SDR} are strongly connected:

\begin{proposition}\label{prop: main}
 Suppose that problem \eqref{WSP-SDR} is feasible. Then problems \eqref{WPP-SDR} and \eqref{WSP-SDR} attain the same optimal objective value. Moreover, if $(\Wb_1^\star,\ldots,\Wb_K^\star,\lambda_1^\star,\ldots,\lambda_K^\star)$ is an optimal solution of problem \eqref{WSP-SDR}, then there exists $(\Vb_1^\star,\ldots,\Vb_K^\star)$ such that $(\Vb_1^\star,\ldots,\Vb_K^\star,\Wb_1^\star,\ldots,\Wb_K^\star)$ is an outer-inner solution of problem \eqref{WPP-SDR}.
\end{proposition}

As the main contribution of this paper, Proposition 1 provides a solution correspondence between problems \eqref{WPP-SDR} and \eqref{WSP-SDR}, showing
that problem \eqref{WPP-SDR} is actually a dual representation of problem \eqref{WSP-SDR}.
To prove that problems \eqref{WPP-SDR} and \eqref{WSP-SDR} attain the same optimal objective value, we show that the Lagrangian dual of problem \eqref{WSP-SDR} is equivalent to the Lagrangian dual of problem \eqref{WPP-SDR}. The former can be shown to be
\begin{align}\label{wsp-sdr2}
\max_{\substack{\Ab_i \in \mathbb{H}^{N_t+1},\\i=1,\ldots,K}}~ &\sum_{i=1}^{K}  \sigma_i^2 [\Ab_i]_{N_t+1} \\
{\rm s.t.}~&
   \Yb_i(\Ab_1,\ldots,\Ab_K)\succeq \zerob,~i=1,\ldots,K, \notag\\
&\tr(\Ab_i) \leq (1+r_i^2) [\Ab_i]_{N_t+1},~i=1,\ldots,K, \notag\\
&\Ab_1,\ldots,\Ab_K \succeq \zerob, \notag
\end{align} where $\Ab_1,\ldots,\Ab_K\in \mathbb{H}^{N_t+1}$ are the (Lagrangian) dual variables associated with constraints \eqref{WSP-SDR C1}, and
%$\Yb_i(\cdot)$ are defined as
\begin{align}
  \label{Yi}
  &\Yb_i(\Ab_1,\ldots,\Ab_K)\triangleq  \Ib- \frac{1}{\gamma_i}\begin{bmatrix}
  \Ib~  \bar\hb_i
\end{bmatrix}\Ab_i\begin{bmatrix}
  \Ib \\ \bar\hb_i^H
\end{bmatrix}  \notag \\
&~~~~~~~~~+\sum_{k=1,k\neq
  i}^{K}\begin{bmatrix}
  \Ib~  \bar\hb_k
\end{bmatrix}\Ab_k\begin{bmatrix}
  \Ib \\ \bar\hb_k^H
\end{bmatrix},~i=1,\ldots,K.
\end{align} Now let us
consider the Lagrangian dual of the inner minimization problem of \eqref{WPP-SDR}, which can be
shown to be
\begin{align}\label{P-D}
\max_{\mu_1,\ldots,\mu_K\geq 0}~&\sum_{i=1}^{K} \mu_i \sigma_i^2 \\
    {\rm s.t.}~&
    \Ib- \frac{\mu_i}{\gamma_i}\Rb_i +\sum_{k=1,k\neq
      i}^{K}\mu_k\Rb_i\succeq \zerob,~i=1,\ldots,K, \notag
\end{align}
where $\mu_1,\ldots,\mu_K$ are the dual variables associated with
the trace inequality constraints of the inner problem of \eqref{WPP-SDR}. Replacing the inner problem
of \eqref{WPP-SDR} with its dual \eqref{P-D}, we obtain the following problem
\begin{align}\label{WPP-SDR inner dual}
\max_{\substack{\Vb_i \in
\mathbb{H}^{N_t+1},\\i=1,\ldots,K}} &\left\{\!\!\!\!\!\!
 \begin{array}{ll}
    &\!\!{\displaystyle \max_{\substack{\mu_i\geq 0,\\i=1,\ldots,K}}~\sum_{i=1}^{K} \mu_i \sigma_i^2} \\
    &~~~~~{\rm s.t.}~
    \Yb_i(\mu_1\Vb_1,\ldots,\mu_K\Vb_K)\succeq \zerob, \\
    &~~~~~~~i=1,\ldots,K,
  \end{array}\!\!\!\!\right\} \\
{\rm s.t.}&~\tr(\Vb_i) \leq (1+r_i^2),~i=1,\ldots,K, \notag \\
         &~[\Vb_i]_{N_t+1}=1,~i=1,\ldots,K,\notag \\
         &~\Vb_1,\ldots,\Vb_K \succeq \zerob. \notag
\end{align}
Since strong duality holds for the inner parts of
\eqref{WPP-SDR} and \eqref{WPP-SDR inner dual}, the two problems
have the same optimal objective value.

One may observe a connection between \eqref{wsp-sdr2} and \eqref{WPP-SDR inner dual}:
\begin{align}
   [\Ab_i]_{N_t+1}=\mu_i,~~\Ab_i  =\mu_i\Vb_i,~i=1,\ldots,K.
\end{align}
In fact, \eqref{wsp-sdr2} and \eqref{WPP-SDR inner dual} are
equivalent problems, as we show in Appendix the following lemma:

\begin{lemma}\label{lemma: (WPP-SDR)=(D)}
 %Problems \eqref{wsp-sdr2} and \eqref{WPP-SDR inner dual} are equivalent.
 If $(\Ab_1^\star,\ldots,\Ab_K^\star)$ is an optimal solution of \eqref{wsp-sdr2}, then
 \begin{align}
 (\Vb_1^\star,\ldots,\Vb_K^\star)&=(\Ab_1^\star/[\Ab_1^\star]_{N_t+1},\ldots,\Ab_K^\star/[\Ab_K^\star]_{N_t+1}),
 \notag \\
 (\mu_1^\star,\ldots,\mu_K^\star)&=([\Ab_1^\star]_{N_t+1},\ldots,[\Ab_K^\star]_{N_t+1})
 \end{align}
 is an optimal outer-inner solution pair of \eqref{WPP-SDR inner dual}. If $(\Vb_1^\star,\ldots,\Vb_K^\star,\mu_1^\star,\ldots,\mu_K^\star)$ is an optimal outer-inner solution of
 \eqref{WPP-SDR inner dual}, then $(\Ab_1^\star,\ldots,\Ab_K^\star)=(\mu_1^\star\Vb_1^\star ,\ldots,\mu_K^\star\Vb_K^\star)$ is optimal to \eqref{wsp-sdr2}.
\end{lemma}
Lemma \ref{lemma: (WPP-SDR)=(D)} shows that
$(\Vb_1^\star,\ldots,\Vb_K^\star)$ of \eqref{WPP-SDR inner dual} only differs
from $(\Ab_1^\star,\ldots,\Ab_K^\star)$ of \eqref{wsp-sdr2} up to a
positive scalar. Hence, \eqref{WPP-SDR inner dual} and  \eqref{wsp-sdr2} attain the same optimal objective value, implying that \eqref{WPP-SDR} and \eqref{WSP-SDR} attain the same optimal objective value. By Lemma 1, one can further show that  $(\Wb_1^\star,\ldots,\Wb_K^\star)$, the optimal primal solution of \eqref{WSP-SDR}, is also optimal to \eqref{WPP-SDR}. The detailed proof is presented in Appendix.

\subsection{Implication and Concluding Remark}

To show that the robust SDR problem \eqref{WSP-SDR} has a rank-one solution, we still need to prove that the optimal $(\Wb_1,\ldots,\Wb_K)$ of \eqref{WPP-SDR} is also optimal to \eqref{WSP-SDR}. Now, let us
%make a conjecture:
assume:
\begin{Condition}
The optimal solution of the inner minimization of problem \eqref{WPP-SDR},
$(\Wb_1^\star,\ldots,\Wb_K^\star)$, is unique.
% The inner minimization of problem \eqref{WPP-SDR} has a unique optimal $(\Wb_1^\star,\ldots,\Wb_K^\star)$.
\end{Condition}
Condition 1 is considered mild; by numerical experience, Condition 1 is found to hold all the time.
Under Condition 1, we can infer from Fact 1 and Proposition 1 that the SDR problem \eqref{WSP-SDR} has a rank-one solution.
%, thus
%establishing the optimality of the SDR problem \eqref{WSP-SDR} for the original worst-case robust beamforming problem \eqref{eq. robust beamforming}.
%
Hence, we conclude that
\begin{Claim}
Under Condition~1, the SDR problem \eqref{WSP-SDR} solves the worst-cast robust problem \eqref{eq. robust beamforming} optimally.
\end{Claim}

%While a rigorous proof for Conjecture 1 is not available yet, Conjecture 1 may be justified as follows. Suppose that \eqref{PSDR} has multiple rank-one solutions. Then \eqref{PSDR} must have higher rank solutions since they can be obtained by convex combination of rank-one solutions. Besides, it is observed numerically that the interior-point algorithms \cite{BK:BoydV04}, which use the logarithmic determinant barrier term to handle the PSD constraints, will in general converge to the higher rank solution, if it exits, due to the fact that the higher rank solution has less penalty on the logarithmic determinant barrier term. Since the simulation experiments keep showing that the interior-point algorithms, when applied to solve \eqref{PSDR}, yield rank-one solutions\footnote{In our simulation experiments, we used the interior-point method based solver \text{SeDuMi} \cite{sedumi} to handle problem \eqref{PSDR}.}, it follows from the above facts that problem \eqref{PSDR} should have a unique rank-one solution for the tested problem instances.

Our analysis above narrows down the SDR optimality question to the proof of unique rank-one solution of the inner minimization problem of \eqref{WPP-SDR}.
As a future research direction, it would be interesting to investigate sufficient conditions under which Condition 1 holds true.

%%%%%%%%%%%%%%%%%%%%%%%%%%%%%%%%%%%%%%%%%%%%%%%%%%%%%%%%%%%%%%%%%%%%
%%%%%%%%%%%%%%%%%%%%%%%%%%%%%%%%%%%%%%%%%%%%%%%%%%%%%%%%%%%%%%%%%%%%

\section{Appendix}

\noindent \underline{{\bf KKT conditions of \eqref{WSP-SDR}}} \\
The KKT conditions of \eqref{WSP-SDR} and \eqref{wsp-sdr2} can be shown
to be
\begin{subequations}\label{KKT of RPM-SDR}
\begin{align}
\!\!\!\!\!&\Wb_1,\ldots,\Wb_K\succeq
  \zerob,\lambda_1,\ldots,\lambda_K \geq 0,\Ab_1,\ldots,\Ab_K\!\!\succeq
  \zerob, \label{primal feasible}\\
& \Psib_i(\Wb_1,\ldots,\Wb_K,\lambda_i) \succeq
\zerob,~i=1,\ldots,K, \label{Psib_i PSD}\\
&\Yb_i(\Ab_1,\ldots,\Ab_K)\succeq \zerob,~i=1,\ldots,K, \label{T_i
PSD}
\\
&\Psib_i(\Wb_1,\ldots,\Wb_K,\lambda_i)\Ab_i =\zerob,~i=1,\ldots,K, \label{complementary condition A}\\
&\Yb_i(\Ab_1,\ldots,\Ab_K)\Wb_i=\zerob,~i=1,\ldots,K, \label{complementary condition}\\
& \tr(\Ab_i) \leq (1+r_i^2) [\Ab_i]_{N_t+1},~i=1,\ldots,K, \label{dual feasible}\\
&\!\left(\tr(\Ab_i)-(1+r_i^2)[\Ab_i]_{N_t+1}\right)\lambda_i=0,~i=1,\ldots,K,
\label{dual feasible2}
\end{align}
\end{subequations}
where $\Psib_i(\cdot)$ and $\Yb_i(\cdot)$ are defined in
\eqref{definition of Psib_i} and \eqref{Yi}, respectively.

\noindent \underline{\bf{Proof of Lemma \ref{lemma: (WPP-SDR)=(D)}:}} Lemma \ref{lemma: (WPP-SDR)=(D)} can be
easily proved by inspection of \eqref{WPP-SDR inner
dual} and \eqref{wsp-sdr2}. What remains is to show that
$\mu_i^\star>0$ and $[\Ab_i^\star]_{N_t+1}>0$ for all
$i=1,\ldots,K$. The former has been proved in \cite[Proposition
4.2]{Huang2010-2}; while the latter can be proved as follows.
One can observe from \eqref{primal feasible} and \eqref{dual
feasible} that $[\Ab_i^\star]_{N_t+1} =0$ results in
$\Ab_i^\star=\zerob$. In this
case, $\Yb_i(\Ab_1^\star,\ldots,\Ab_K^\star)$ in \eqref{Yi} is
positive definite, i.e., $\Yb_i(\Ab_1^\star,\ldots,\Ab_K^\star)
\succ \zerob.$ By the complementary slackness \eqref{complementary
condition}, this leads to the primal solution $\Wb_i^\star=\zerob$,
which however violates \eqref{Psib_i PSD} [see \eqref{definition of
Psib_i}] due to $\sigma_i^2>0$. \hfill $\blacksquare$

\noindent \underline{\bf{Proof of Proposition \ref{prop: main}:}}
Here we prove that $(\Wb_1^\star,\ldots,\Wb_K^\star)$, the optimal primal solution of \eqref{WSP-SDR}, is also optimal to \eqref{WPP-SDR}.
By Lemma \ref{lemma: (WPP-SDR)=(D)} which shows that
$(\Ab_1^\star/[\Ab_1^\star]_{N_t+1},$ $\ldots,\Ab_K^\star/[\Ab_K^\star]_{N_t+1})$
is an optimal outer maximizer of \eqref{WPP-SDR}, it suffices
to show that $(\Wb_1^\star,\ldots,\Wb_K^\star)$ is optimal to the following problem
\begin{align}\label{inner problem4}
\min_{\Wb_1,\ldots,\Wb_K \succeq \zerob}
~ &\sum_{i=1}^{K} {\tr}(\Wb_i)\\
{\rm s.t.}~&
 \tr\left(\left(\frac{1}{\gamma_i}\Wb_i - \sum_{k=1,k\neq
  i}^{K}\Wb_k\right)\begin{bmatrix}
  \Ib~  \bar\hb_i
\end{bmatrix}\Ab_i^\star\begin{bmatrix}
      \Ib\\  \bar\hb_i^H
    \end{bmatrix}\right)  \notag \\
    &\geq
\sigma_i^2[\Ab_i^\star]_{N_t+1},~i=1,\ldots,K.\notag
\end{align}
This can be shown by examining that $(\Wb_1^\star,\ldots,\Wb_K^\star)$ satisfies the KKT conditions of \eqref{inner problem4}, which are given as follows:
\begin{subequations}\label{inner problem4 KKT}
\begin{align}
&\Wb_1,\ldots,\Wb_K\succeq
  \zerob,~\mu_1,\ldots,\mu_K\geq
0,  \label{inner problem4 K1} \\
&\Yb_i\left(\mu_1(\Ab_1^\star/[\Ab_1^\star]_{N_t+1}),\ldots,{\mu_K}(\Ab_K^\star/{[\Ab_K^\star]_{N_t+1}})\right)\succeq \zerob, \label{inner problem4 K2}\\
&\Yb_i\left(\mu_1(\Ab_1^\star/[\Ab_1^\star]_{N_t+1}),\ldots,{\mu_K}(\Ab_K^\star/
{[\Ab_K^\star]_{N_t+1}})\right)\Wb_i=\zerob,\label{inner problem4 K3}\\
&\tr\left(\left(\frac{1}{\gamma_i}\Wb_i - \!\!\!\!\sum_{k=1,k\neq
  i}^{K}\!\!\!\Wb_k\right)\begin{bmatrix}
  \Ib~  \bar\hb_i
\end{bmatrix}\Ab_i^\star\begin{bmatrix}
      \Ib\\  \bar\hb_i^H
    \end{bmatrix}\right)\!\!\! = \sigma_i^2[\Ab_i^\star]_{N_t+1}, \label{inner problem4 K4}
\end{align}
\end{subequations} for $i=1,\ldots,K$.
%We will show that $(\Wb_1^\star,\ldots,\Wb_K^\star)$ satisfies all
%the conditions in \eqref{inner problem4 KKT}.

Since
$(\Wb_1^\star,\ldots,\Wb_K^\star,\lambda_1^\star,\ldots,\lambda_K^\star)$
and $(\Ab_1^\star,\ldots,\Ab_K^\star)$ satisfy the KKT conditions in
\eqref{primal feasible}, \eqref{T_i PSD} and \eqref{complementary
condition}, $(\Wb_1^\star,\ldots,\Wb_K^\star)$ and
$(\mu_1,\ldots,\mu_K)\triangleq
([\Ab_1^\star]_{N_t+1},\ldots,[\Ab_K^\star]_{N_t+1})$ satisfy
\eqref{inner problem4 K1}, \eqref{inner problem4 K2} and
\eqref{inner problem4 K3}. To show that
$(\Wb_1^\star,\ldots,\Wb_K^\star)$ also fulfills \eqref{inner
problem4 K4}, let us consider an alternative representation of
\eqref{WSP-SDR}:
\begin{lemma} Problem \eqref{WSP-SDR} can be equivalently expressed as the
following problem
\begin{subequations} \label{WSP SDR2}
\begin{align}
\min_{\substack{\Wb_i\succeq \zerob,\\i=1,\ldots,K}}~&
\sum_{i=1}^{K} {\tr}(\Wb_i) \label{obj2}\\
{\rm s.t.} ~ &\min_{\Vb_i \in \Vc_i}\!\! \tr
   \left(\left(\frac{1}{\gamma_i}\Wb_i-\!\!\!\sum_{k=1,k\neq
  i}^{K}\Wb_k\right)\begin{bmatrix}
  \Ib~  \bar\hb_i
\end{bmatrix}   \Vb_i\begin{bmatrix}
      \Ib\\  \bar\hb_i^H
    \end{bmatrix}\right) \notag \\
    &~~~~~~~\geq \sigma_i^2,~i=1,\ldots,K, \label{subproblem3}
\end{align}
\end{subequations} where $
\Vc_i=\{\Vb_i\in \mathbb{H}^{N_t+1}~|~\tr(\Vb_i) \leq (1+r_i^2),
[\Vb_i]_{N_t+1}=1, \Vb_i\succeq
   \zerob\}.
$
\end{lemma}
It is easy to verity that, for
$(\Wb_1^\star,\ldots,\Wb_K^\star)$,
\begin{align}
\min_{\Vb_i \in \Vc_i}~ &\tr
   \left(\left(\frac{1}{\gamma_i}\Wb_i^\star-\sum_{k=1,k\neq
  i}^{K}\Wb_k^\star\right)\begin{bmatrix}
  \Ib~  \bar\hb_i
\end{bmatrix}   \Vb_i\begin{bmatrix}
  \Ib \\ \bar\hb_i^H
\end{bmatrix}\right) \notag \\
&= \sigma_i^2,~i=1,\ldots,K,
\end{align} i.e., the inequality constraints in \eqref{subproblem3} are all
active for the optimal solution $(\Wb_1^\star,\ldots,\Wb_K^\star)$.
Hence, to show that \eqref{inner problem4 K4} is also fulfilled by
$(\Wb_1^\star,\ldots,\Wb_K^\star)$, it is sufficient to prove that
\begin{subequations}\label{error prob SDR2}
\begin{align}
&\Ab_i^\star/[\Ab_i^\star]_{N_t+1} = \notag \\
&\arg~\min_{\Vb_i \succeq
\zerob}~ \tr
   \left(\left(\frac{1}{\gamma_i}\Wb_i^\star-\sum_{k=1,k\neq
  i}^{K}\Wb_k^\star\right)\begin{bmatrix}
  \Ib~  \bar\hb_i
\end{bmatrix}   \Vb_i\begin{bmatrix}
  \Ib \\ \bar\hb_i^H
\end{bmatrix}\right) \notag \\
&~~~~~~~~~~~{\rm s.t.}~\tr(\Vb_i) \leq (1+r_i^2),  \label{subproblem4 C1}\\
&~~~~~~~~~~~~~~~~[\Vb_i]_{N_t+1}=1, \label{subproblem4 C2}
\end{align}
\end{subequations} for all $i=1,\ldots,K$. Let $\xi_i$ and $\tau_i$
be the dual variables associated with the constraints in
\eqref{subproblem4 C1} and \eqref{subproblem4 C2}, respectively, and
define
\begin{align*} %\label{hat Psib_i}
&\tilde\Psib_i\left(\Wb_1^\star,\ldots,\Wb_K^\star,\xi_i,\tau_i\right)& \notag \\&\triangleq
\begin{bmatrix}
\Ib \\
\bar \hb_i^H
\end{bmatrix}\!\!\left(\frac{1}{\gamma_i}\Wb_i^\star-\!\!\!\sum_{k=1,k\neq
i}^{K}\Wb_k^\star\right)\!\!
\begin{bmatrix}
\Ib ~\bar \hb_i
\end{bmatrix}\! +\!
\begin{bmatrix} \xi_i \Ib & {\bf 0}\\
{\bf 0} & \xi_i+\tau_i
\end{bmatrix}.
\end{align*}
The KKT conditions of the minimization problem in \eqref{error prob
SDR2} can be obtained as
\begin{subequations}\label{KKT of subproblem}
\begin{align}
\!\!\!\!\!\!\!\!\!\!\!\!\!\!\!\!\!\!\!\!\!\!&\tr(\Vb_i)\!\! \leq\!\! (1+r_i^2),[\Vb_i]_{N_t+1}\!=\!1, \!\Vb_i\! \succeq\!
\zerob, \xi \geq 0,\tau_i \in \mathbb{R}, \\
&
\tilde\Psib_i\left(\Wb_1^\star,\ldots,\Wb_K^\star,\xi_i,\tau_i\right)
\succeq \zerob, \\
&
\tilde\Psib_i\left(\Wb_1^\star,\ldots,\Wb_K^\star,\xi_i,\tau_i\right)\Vb_i
=\zerob, \\
& \xi_i\left(\tr(\Vb_i)-(1+r_i^2)\right)=0,~
\tau_i\left([\Vb_i]_{N_t+1}-1\right)=0.
\end{align}
\end{subequations}
For each $i\in \{1,\ldots,K\}$, let
$
  \Vb_i^\star=\Ab_i^\star/[\Ab_i^\star]_{N_t+1},~
  \xi_i^\star=\lambda_i^\star,~\tau_i^\star=-\sigma_i^2-(1+r_i^2)\lambda_i^\star.
$ It follows from the KKT conditions in \eqref{primal feasible},
\eqref{Psib_i PSD}, \eqref{complementary condition A} and
\eqref{dual feasible2} that $(\Vb_i^\star,\xi_i^\star,\tau_i^\star)$
satisfies all the conditions in \eqref{KKT of subproblem}. Thus
\eqref{error prob SDR2} is true for all $i=1,\ldots,K$. The proof is
then completed. \hfill $\blacksquare$

\noindent \underline{\bf{Proof of Lemma 2:}}
It suffices to show that \eqref{WSP-SDR C1} is equivalent
to \eqref{subproblem3}. Note that
\eqref{WSP-SDR C1} is equivalent to
\begin{align} \label{subproblem}
\min_{\|\eb_i\|^2 \leq r_i^2} &\left\{ (\bar\hb_i+\eb_i)^H
\left(\frac{1}{\gamma_i}\Wb_i-\sum_{k=1, k \neq
  i}^{K}\Wb_k\right)(\bar\hb_i+\eb_i) \right\} \notag \\
  & \geq \sigma_i^2,~i=1,\ldots,K.
\end{align} (the equivalence is owing to the $\mathcal{S}$-Lemma; see  \cite{ZhengWongNg_2008,Song2011}).
Note that the minimization problem on the left-hand side of
\eqref{subproblem} may not be convex with respect to
$(\eb_1,\ldots,\eb_K)$ because the matrix
$(\frac{1}{\gamma_i}\Wb_i-\sum_{ k \neq i}^{K}\Wb_k)$ may not be
positive semidefinite. Nevertheless, SDR can be applied. Through the same procedure as in obtaining \eqref{WPP-SDR}, one can obtain the SDR
problem of the minimization problem in \eqref{subproblem} as
\begin{align}\label{error prob SDR}
\!\!\!\!   \min_{\Vb_i \in \Vc_i}~& \tr
   \left(\left(\frac{1}{\gamma_i}\Wb_i-\sum_{k=1,k\neq
  i}^{K}\Wb_k\right)\begin{bmatrix}
  \Ib~  \bar\hb_i
\end{bmatrix}   \Vb_i\begin{bmatrix}
  \Ib \\ \bar\hb_i^H
\end{bmatrix}\right).
\end{align}  While \eqref{error prob SDR} is obtained by
relaxation of the rank of $\Vb_i$, the SDR problem \eqref{error prob
SDR} is actually tight and optimal to the minimization problem in
\eqref{subproblem}; see
\cite[Lemma 3.1]{Huang09}. We thus obtain
\eqref{subproblem3} by substituting \eqref{error prob SDR} into
\eqref{subproblem}. \hfill $\blacksquare$

\section{Acknowledgements}
This work is supported in part by National Science Council,
R.O.C., under Grant NSC-99-2221-E-007-052-MY3, 
by a General Research Fund of Hong Kong Research Grant Council (CUHK 415908),
and by a Direct Grant awarded by the Chinese
University of Hong Kong (Project Code 2050489).

\vspace{-0.0cm} \footnotesize
\bibliography{refs10}
\end{document}